\documentclass[twoside]{article}

\usepackage[accepted]{aistats2021}
\usepackage{amsfonts, amsmath, amssymb, amsthm, bbm, wasysym}
\usepackage{mathrsfs, mathtools}
\usepackage{dsfont}
\usepackage{wrapfig}
\usepackage{xcolor}
\usepackage{comment}
\usepackage{rotating}
\usepackage{algorithmic}
\usepackage[algo2e,ruled,linesnumbered,shortend,lined,vlined,noend]{algorithm2e}
\usepackage[all]{xy}
\usepackage[capitalize]{cleveref}
\usepackage{arydshln}
\usepackage{enumitem}
\usepackage{graphicx}
\usepackage{multirow}
\usepackage{booktabs}
\usepackage{slantsc}
\usepackage{url}
\usepackage{wrapfig}  
\newtheorem{theorem}{Theorem}
\DeclarePairedDelimiter\floor{\lfloor}{\rfloor}

\newcommand{\cb}{\mathbf{c}}

\begin{document}

%
\runningtitle{ATOL: Measure Vectorization for Automatic Topologically-Oriented Learning}

%

\twocolumn[

\aistatstitle{ATOL: Measure Vectorization for \\ Automatic Topologically-Oriented Learning}

\aistatsauthor{ Martin Royer \And Fr\'ed\'eric Chazal}

\aistatsaddress{ 	Datashape, Inria Saclay, Palaiseau, France. }

\aistatsauthor{ Cl\'ement Levrard }

\aistatsaddress{ LPSM, Univ. Paris Diderot, Paris, France. }

\aistatsauthor{ Umeda Yuhei \And  Ike Yuichi }

\aistatsaddress{ Fujitsu Laboratories, AI Lab, Tokyo, Japan. } ]

\begin{abstract}
	Robust topological information commonly comes in the form of a set of persistence diagrams, finite measures that are in nature uneasy to affix to generic machine learning frameworks. We introduce a fast, learnt, unsupervised vectorization method for measures in Euclidean spaces and use it for reflecting underlying changes in topological behaviour in machine learning contexts. The algorithm is simple and efficiently discriminates important space regions where meaningful differences to the mean measure arise. It is proven to be able to separate clusters of persistence diagrams. We showcase the strength and robustness of our approach on a number of applications, from emulous and modern graph collections where the method reaches state-of-the-art performance to a geometric synthetic dynamical orbits problem. The proposed methodology comes with a single high level tuning parameter: the total measure encoding budget. We provide a completely open access software.
\end{abstract}

\section{Introduction}
\label{sec:intro}

Topological Data Analysis (TDA) is a field dedicated to the capture and description of relevant geometric or topological information from data. The use of TDA with standard machine learning tools has proved particularly advantageous in dealing with all sorts of complex data, meaning objects that are not or only partly Euclidean, for instance graphs, time series, etc. The applications are abundant, from social network analysis, bio and chemoinformatics, to physics, imaging and computer vision, to name a few. Recent examples include \cite{dindin20}, \cite{pike20}, \cite{duponchel18}, \cite{cole19}, \cite{KHASAWNEH18}. Through Persistent Homology, a multi-scale analysis of the topological properties of the data, robust information is extracted. But the resulting features are commonly generated in the form of a persistence diagram whose structure does not easily fit the general machine learning input format. So TDA captures relevant information in a form that is challenging to handle \--- therefore it is generally combined to machine learning by way of an embedding method for persistence diagrams. This work is set in that trend. 

\noindent
\textbf{Contributions}. First we introduce a learnt, unsupervised vectorization method for measures in Euclidean spaces of any dimension (Section \ref{sec:measure-vect}). Then we show how this method can be used for Topologically-Oriented Learning (Section \ref{sec:topological-learning}), allowing for easy integration of topological features such as persistence diagrams into challenging machine learning problems. We illustrate our approach with sets of experiments that lead to state-of-the-art results on challenging problems (Section \ref{sec:applications}). We provide an open source implementation.

Our algorithm is simple and easy to use. It relies on a quantization of the space of diagrams that is statistically optimal. It is fast and practical for large scale and high dimensional problems. It is competitive and sometimes largely surpasses more sophisticated methods involving kernels, deep learning, or computations of Wasserstein distance. To the best of our knowledge, we introduce the first vectorization method for persistence diagrams that is proven to be able to separate clusters of persistence diagrams. There is little to no tuning to this method, and no knowledge of TDA is required for using it.

\noindent
\textbf{Related work}. Finding representations of persistence diagrams that are well-suited to be combined with standard machine learning pipeline is a problem that has attracted a lot of interest these last years.  
A first family of approaches consists in finding convenient vector representations of persistence diagrams. For instance it involves interpreting diagrams as images in \cite{Adams2017}, extracting topological signatures with respect to fixed points whose optimal position are supervisedly learnt in \cite{hofer2017deep}, a square-root transform of their approximated pdf in \cite{Anirudh}. Recently \cite{Perea2019ApproximatingCF} introduced template functions, a mathematical framework to understand featurisation functions that integrates against the measure of a persistence diagram; our method is interpretable in this framework.
A second family of approaches consists in designing specific kernel on the space of persistence diagrams, such as the multi-scale kernel of \cite{Reininghaus2015}, the weighted Gaussian kernel of \cite{Kusano2016} or the sliced Wasserstein kernel of \cite{Carriere2017}. Those techniques have state-of-the-art behaviour on problems, but for drawback they require another step for an explicit representation, and are known to scale poorly.
A recent other line of work has managed to directly combine the uneasy structure of persistence diagrams to neural networks architectures \cite{Zaheer2017}, \cite{perslay}. Despite their successful performances, these neural networks are heavy to deploy and hard to understand. They are sometimes paired with a representation method as in \cite{hofer2017deep}, \cite{hofer2019graph}.

\begin{figure}
	\centering
	\includegraphics[width=.3 \columnwidth]{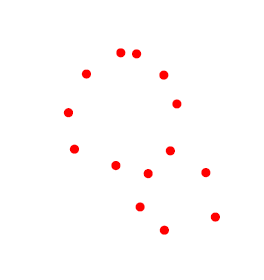}
	\includegraphics[width=.3 \columnwidth]{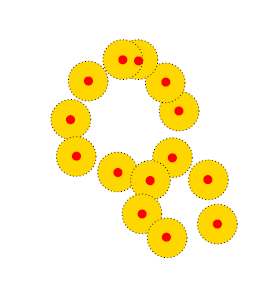}
	\includegraphics[width=.3 \columnwidth]{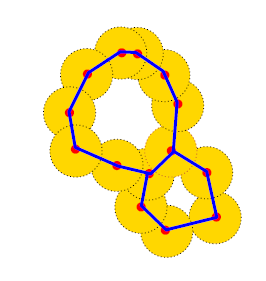}\\
	\includegraphics[width=.3 \columnwidth]{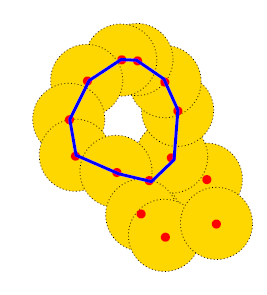}
	\includegraphics[width=.3 \columnwidth]{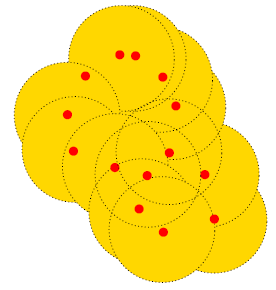}
	\includegraphics[width=.3 \columnwidth]{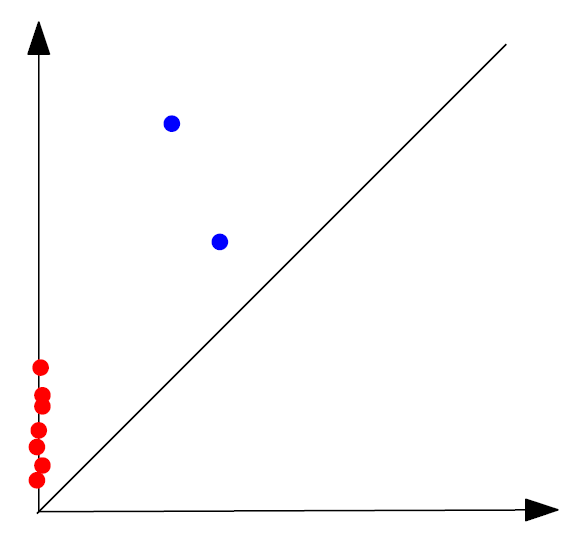}
	\caption{filtration by union of balls built on top of a 2-dimensional data set (red points) and its corresponding persistence diagram. As the balls radii increase (from left to right and top to bottom), the connected components (red points) are merged; two-cycles (blue points) appear and disappear along the filtration.}
	\label{fig:persistence}
\end{figure}

\textbf{Persistent homology in TDA}. Persistent homology provides a rigorous mathematical framework and efficient
algorithms to encode relevant multi-scale topological features of complex data such as point clouds, time-series, 3D images... More precisely, persistent homology encodes the evolution of the topology of families of nested topological spaces $(F_\alpha)_{\alpha \in A}$, called filtrations,  built on top of the data and indexed by a set of real numbers $A$ that can be seen as scale parameters. For example, for a point cloud in a Euclidean space, $F_\alpha$ can be the union of the balls of radius $\alpha$ centered on the data points - see Figure \ref{fig:persistence}.  Given a filtration $(F_\alpha)_{\alpha \in A}$, its topology (homology) changes as $\alpha$ increases: new connected components
can appear, existing connected components can merge, loops and cavities can appear or be filled, etc. Persistent homology tracks these changes, identifies features and associates, to each of them, an interval or lifetime from $\alpha_{birth}$ to $\alpha_{death}$.
For instance, a connected component is a feature that is born at the smallest $\alpha$ such that the component is present in $F_\alpha$, and dies when it merges with an older connected component. The set of intervals representing the lifetime of the identified features
is called the barcode of the filtration. As an interval can also be represented as a point in the plane with coordinates $(\alpha_{birth}, \alpha_{death})$, the persistence barcode is equivalently represented as an union of such points and called the persistence diagram - see \cite{edelsbrunner2010computational,boissonnat2018geometric} for a more detailed introduction.

The classical main advantage of persistence diagrams is that: {\em (i)} they are proven to provide robust qualitative and quantitative topological information about the data \cite{chazal2012structure}; {\em (ii)} since each point of the diagram represents a specific topological feature with its lifespan, they are easily interpretable as features; {\em (iii)} from a practical perspective, persistence diagrams can be efficiently computed from a wide family of filtrations \cite{gudhi}.
However, as persistence diagrams come as unordered  set of points with non constant cardinality, they cannot be immediately processed as standard vector features in machine learning algorithms. Considering diagrams as measures has proven beneficial in the literature before (see for instance \cite{chazal2012structure}, \cite{chazal2018}) and allows to naturally encode the points multiplicity problems in the form of weighted measures. 

\textbf{Notations}. Consider $\mathcal{M}_d$ the set of finite measures on the $d$-dimensional ball $\mathcal{B}_d(0,R)$ of the Euclidean space $\mathbb{R}^d$ with total mass smaller than $M$, for some given $M,R \in \mathbb{R}_+^2$. For $m \in \mathcal{M}_d$ and $\chi: \mathbb{R}^d \rightarrow \mathbb{R}$ borelian function, let $\chi \boldsymbol{\cdot} m := \int_{x \in \mathbb{R}^d} \chi(x) m(dx)$ whenever $|\chi| \boldsymbol{\cdot} m$ is finite.

Next, for $b \in \mathbb{N}^*$ we call a codebook $\cb = (c_1, \dots, c_b) \in \mathcal{B}_d(0, R)^b$ the support of a distribution supported on $b$ points and its associated Voronoï cells: $W_j(\cb) = \{ x \in \mathbb{R}^d \mid \forall i < j, \| x - c_j\| < \| x-c_i\| \text{ and } \forall{i >j}, \| x- c_j\| \leqslant \|x-c_i\| \}$.

Finally, we assume that the set of input persistence diagrams comes as an i.i.d. sample from a distribution of uniformly bounded diagrams, that is given $M,R \in \mathbb{R}_+^2$, let $\mathcal{D}$ be the space of persistence diagrams with at most $M$ points contained in the Euclidean disc $\mathcal{B}_2(0,R)$. The space $\mathcal{D}$ is considered as a subspace of the set $\mathcal{M}_2$ of finite measures on $\mathcal{B}_2(0,R)$  with total mass smaller than $M$: for any $D \in \mathcal{D}$, $D := \sum_{p \in D} \delta_p$ where $\delta_p$ is the Dirac measure centered at point $p$.

\section{Methodology}

In this section we introduce \textsc{Atol}, a simple unsupervised data-driven method for measure vectorization. \textsc{Atol} allows to automatically convert a distribution of persistence diagrams into a distribution of feature vectors that are well-suited for use as topological features in standard machine learning pipelines.

As an overview, given a positive integer $b$, \textsc{Atol} proceeds in two steps: it computes a discrete measure in $\mathbb{R}^d$ supported on $b$ points that approximates the average measure of the distribution from which the input observations have been sampled. Then, it computes a set of well-chosen contrast functions centered on each point of the support of this measure, that are used to convert each observed measure into a vector of size $b$. This resulting vectorization can then be used in standard machine-learning problems such as clustering, classification, etc.

\subsection{Measure vectorization through quantization} \label{sec:measure-vect}

We now introduce Algorithm~\ref{alg:atolfeat} \textsc{Atol}-featurisation: a featurisation method for elements of $\mathcal{M}_d$. The first step in our procedure is to use quantization in space $\mathcal{M}_d$. Starting from an i.i.d. sample of measures $X_1,\dots, X_n$ drawn from probability distribution $\mathcal{L}_X$ on $\mathcal{M}_d$ and given an integer budget $b\in\mathbb{N}^*$, we produce a compact representation for the mean measure $\mathbb{E}(X)$. That is, we produce a distribution $P_{\hat{\cb}_n}$ supported on a fixed-length codebook $\hat{\cb}_n = (c_1, \dots, c_b) \in \mathcal{B}_d(0, R)^b$ that aims to minimize over such distributions $P$ based on $b$ points the distorsion $R(P) := W_2^2(P, \mathbb{E}(X))$: the squared 2-Wasserstein distance to the mean measure. In practice, one considers the empirical mean measure $\bar{X}_n$ and the $k$-means problem for this $\bar{X}_n$ measure. Then the adaptation of Lloyd's \cite{Lloyd82} algorithm to measures can be used.

From this quantization our aim is to derive spatial information on measures in order to discriminate them. Much like one would compactly describe a point cloud with respect to its barycenter in a PCA procedure, we describe measures based on a number of reduced difference to our mean measure approximate. To this end, our second step is to tailor $b$ individual contrast functions each based on the estimated codebook that individually describe the space with respect to a certain viewpoint. In other words we set to find regions of the space where measures seem to aggregate on average, and build a dedicated descriptor for those regions. We define and use the following contrast family $\mathbb{R}^d \rightarrow \mathbb{R}_+$, for $i\in [b]$:
\begin{align}
\label{contrast}
	\Psi_i(\cdot, \hat{\cb}_n) : x \mapsto \exp\big[-\frac{\|x - c_i \|_2}{\sigma_i(\hat{\cb}_n)} \big]
\end{align}
where
\begin{align}
	\label{spread} \sigma_i(\hat{\cb}_n) := \min_{j\in [b], j\neq i} \|c_i - c_j\|_2/2.
\end{align}

These specific contrast functions are chosen to decrease away from the approximate mean centroid in a Laplacian fashion and we choose the scale to correspond to the minimum distance to the closest Voronoi cell in the corresponding codebook $\hat{\cb}_n$.
To our knowledge there is nothing that prevents other, well designed contrast families to be substituted in their place, and to that regard some intuition is provided in the ablation study of Section \ref{sec:ablation}.

Given a mean measure codebook approximate $\hat{\cb}_n$, element $X \in \mathcal{M}_d$ can now be compactly described through the integrated contribution to each contrast functions: $\Psi_i(\cdot, \hat{c}_{n,b}) \boldsymbol{\cdot} X$. Our algorithm concatenates into a vector each of those contributions.

\begin{algorithm2e} 
	\SetAlgoLined
	\KwData{Collection of measures $X_1,\dots, X_n \in (\mathcal{M}_d)^n$.} \SetKwInOut{Input}{Parameters} \Input{budget $b \in \mathbb{N}^*$.}
	Quantization algorithm of the mean measure with fixed-length support (Lloyd's adaptation to measures): sample $\cb = (c_1, \dots, c_b)$ from $\bar{X}_n$. \;
	\While{$\cb^{\textit{new}} = (c_1^{\textit{new}}, \dots, c_b^{\textit{new}}) \neq \cb$}
	{
		$\cb = \cb^{\textit{new}}$ \;
		$\forall i \in [b], \quad
		c^{\textit{new}}_i := \left[ u \mapsto u 1_{W_i(\cb)}(u) \right] \boldsymbol{\cdot} \dfrac{1}{\bar{X}_n (W_i(\cb))} \bar{X}_n$ \;
	}
	let $\hat{\cb}_n$ be the resulting codebook, define the $b$ measurable contrast functions $(\Psi_1(\cdot, \hat{\cb}_n), \dots, \Psi_b(\cdot, \hat{\cb}_n))$ to compute featurisation map: $v_\textsc{Atol}: X \mapsto \Big[ \Psi_i(\cdot, \hat{\cb}_n) \boldsymbol{\cdot} X \Big]_{i \in [b]}$. \;
	\KwResult{vectorization map $v_\textsc{Atol}: \mathcal{M}_d \rightarrow \mathbb{R}^b$.}
	\caption{\textsc{Atol}-vectorization}\label{alg:atolfeat}
\end{algorithm2e}


This algorithm is practical for large scale and high dimensional problems: it has a running time in $O(n \times M \times b \times d)$, so therefore it is able to handle difficult problems as long as corresponding measures are found. If necessary, a single-pass quantization step can be derived as a minibatch adaptation of the \cite{MacQueen67} MacQueen algorithm (we refer to \cite{levrard20}), and then combined with the contrast functions vectorization. Therefore Algorithm \ref{alg:atolfeat} is simple, fast and automatic once the desired length for vectorization has been chosen. Now let us introduce how it appears in machine-learning contexts.

\subsection{Topological learning with \textsc{Atol}}\label{sec:topological-learning}

Set in the context of a standard learning problem, we introduce Algorithm \ref{alg:atol} \textsc{Atol}: Automatic Topologically-Oriented Learning. Let $\Omega := (X, y)$ with given observations $X$ in some space $\mathcal{X}$ corresponding to a known, partially available or hidden label $y \in \mathcal{Y}$. Assume that one has a way to extract toplogical features from $\mathcal{X}$ (for example a collection of diagrams associated to those elements), and let $\kappa: \mathcal{X} \rightarrow \mathcal{M}_d$ be the corresponding map. Then applying Algorithm \ref{alg:atolfeat} to the resulting collection of descriptors provides some simplified topological understanding on elements $X$ of this problem.

\begin{algorithm2e} 
	\SetAlgoLined
	\KwData{Learning problem $\Omega := (X, y)$ with $X \in \mathcal{X}$ collections and $y \in \mathcal{Y}$ labels.}	\SetKwInOut{Input}{Parameters}
	\Input{$\kappa: \mathcal{X} \rightarrow \mathcal{M}_d$ yielding topological descriptors, and budget $b \in \mathbb{N}^*$.}
	Compute intermediate learning problem ${\Omega}_{\text{Topo}} := ((X, \kappa(X)), y) \in (\mathcal{X} \times \mathcal{M}_d) \times \mathcal{Y}$ with topological features, potentially unfit for general machine learning routines\;
	Use Algorithm \ref{alg:atolfeat} to derive Euclidean representations of those features, i.e. transform ${\Omega}_{\text{Topo}}$ into a generic machine learning problem $\widetilde{\Omega} := ((X, v_\textsc{Atol} \circ \kappa(X)), y) \in (\mathcal{X}\times\mathbb{R}^b) \times \mathcal{Y}$. \;
	\KwResult{Enhanced problem $\widetilde{\Omega} := ((X, v_\textsc{Atol} \circ \kappa(X)), y)$ where $v_\textsc{Atol} \circ \kappa(X) \in \mathbb{R}^b$.}
	\caption{\textsc{Atol}: Automatic Topologically-Oriented Learning}\label{alg:atol}
\end{algorithm2e}

This algorithm is integrated in the open source topological library
GUDHI \cite{gudhi} accessible at \url{https://gudhi.inria.fr/python/latest/representations.html}.
We point that as embedding map $v_{\textsc{Atol}}$ is automatically computed without knowledge of a learning task, its derivation is fully unsupervised. The representation is learned since it is data-dependent, but it is also agnostic to the task and eventually only depends on getting a glimpse at an average persistence diagram.

\textbf{\textsc{Atol} in dimension 2 for persistence diagrams.}
We now specialise this algorithm to the context of persistent homology that is usually set in dimension $d = 2$. Applying Algorithm \ref{alg:atol} to a collection from $\mathcal{M}_2$ such as persistence diagrams, as $\mathcal{D} \subset \mathcal{M}_2$, is straightforward and allows to embed the complex, unstructured space $\mathcal{M}_2$ in Euclidean terms.

Now let us assume that the measures in $\mathcal{M}_2$ come from distinct sources: that observed measures $D_1, \dots, D_n$ are sampled with noise from a mixture model $D = \sum_{l=1}^{L} \pi_l D^{(l)}$ of distinct measures $D^{(1)}, \dots, D^{(L)}$ (by that we mean that any two measures in this set differ in support by at least one point). Let $Z$ the latent variable of the mixture so that $D|Z=l \sim D^{(l)}$. The following results ensures that $v_{\textsc{Atol}}$ has separative power, i.e. that the vectorization clearly separates the different sources:
\begin{theorem}[Separation with \textsc{Atol}]\ \\
	\label{thm:separation-atol}
	For a given noise level assuming $\mathbb{E}(D)$ satisfies some (explicit) margin condition and for $n$ and $b$ large enough there exists a non-empty segment for $\sigma_1, \dots, \sigma_b$ in Equation \eqref{contrast} such that for all $i, j \in [n]^2$, with high probability:
	\begin{align}
	Z_i = Z_j &\implies \Vert v_{\textsc{Atol}}(D_i) - v_{\textsc{Atol}}(D_j) \Vert_{\infty} \leqslant 1/4, \\
	Z_i \neq Z_j &\implies \Vert v_{\textsc{Atol}}(D_i) - v_{\textsc{Atol}}(D_j) \Vert_{\infty} \geqslant 1/2.
	\end{align}
\end{theorem}

To our knowledge it is the first time that a measure vectorization method (or a persistence diagram vectorization method) has been proven to separate clusters. This result follows from Corollary 19 in \cite{levrard20} that studies theoretical properties of \textsc{Atol}-like procedures. The explicit statement of the assumptions and margin conditions are standard and rather technical. 

But the theory behind it uses an idealistic framework (including the so-called margin condition) under which such procedures will succeed in separating different sources. Based on this framework, the requirements of Theorem 1 cannot be checked in practice: apart from the technical margin condition, the prescribed bounds on budget $b$ are unknown and they theoretically grow quite large with the number of underlying centers + covering number, and the bounds for bandwidths $\sigma_1, \hdots, \sigma_b$ are heavily dependent on the structure of source model $D$.

In practice we prove it needs not be so difficult: for intuition we refer to the ablation study on the influence of $b$ exposed in Table \ref{tab:study}, that shows it is easy to choose a low budget for efficient results \--- so we leave it as the only parameter of the algorithm. For full automaticity, a simple adaptive strategy would be to try a range of budgets during the training task, since the combination of Algorithm 1 and a standard learning algorithm such as random forests runs very fast. Furthermore, the adaptive strategy of Equation \eqref{spread} for bandwidths $\sigma_1, \hdots, \sigma_b$ proves efficient, see the bandwiths variation study of Section \ref{sec:ablation} Figure \ref{fig:sigmas}.

In dimension 2 this vectorization is conceptually close to two other recent works. \cite{hofer2017deep} computes a persistence diagram vectorization through a deep learning layer that adjusts Gaussian contrast functions used to produce topological signatures. So in essence our approach substitutes quantization to deep learning, with no need of supervision and allowing to provide mathematical guarantees. Next, the bag of word method of \cite{zelinski2019} uses an ad-hoc form of quantization for the space of diagrams, then count functions as contrast functions to produce histograms as topological signatures. Those are in fact sensible differences, that will ultimately translate in terms of effectiveness: Section \ref{sec:graphs} shows the \textsc{Atol}-featurisation to produce state-of-the-art mean accuracy on two difficult multi-class classification problems (67.1 \% on \texttt{REDDIT5K} and 51.4 \% on \texttt{REDDIT12K}) that are also analysed by those papers: \cite{hofer2017deep} report a mean accuracy of respectively 54.5\% and 44.5\%, and \cite{zelinski2019} report an accuracy of respectively 49.9\% and 38.6\%.

\section{Competitive TDA-Learning} \label{sec:applications}

In this section we show the \textsc{Atol} framework to be competitive, sometimes greatly improving the state-of-the-art, but also versatile and easy to use. This section presents experiments on two sorts of classification problems (graphs and point clouds), and another applied experiment on time series is provided in the Supplementary Materials.

Algorithm \ref{alg:atol} transforms the initial problems into a typically standard machine-learning problem, so the problem although transformed remains to be solved. In the instances below we use the \texttt{scikit-learn} \cite{sklearn} random-forest classification tool with $100$ trees and all other parameters set as default. We use random forests as a ready-to-use tool, but comparable performances can be obtained from using a linear SVM classifier or a neural network classifier, depending on the problem. It is a light choice that requires no particular infrastructure or tuning efforts that would produce overly design-dependent results \--- as our ambition is to show an ability to perform well overall.

\subsection{Graph Classification}
\label{sec:graphs}

\begin{table*}
	\resizebox{.9\linewidth}{!}{ 
		\begin{tabular}{|l|rrrrr|r|}
			\hline
			\hspace{1.2cm} method & RetGK & FGSD & WKPI & GNTK & PersLay & \textsc{Atol} \\
			problem & \cite{zhang2018retgk} & \cite{verma2017hunt} & \cite{qi2019} & \cite{du19} & \cite{perslay} & \\
			\hline
			$\texttt{REDDIT}$ (5K, 5 classes)       & 56.1$\pm$.5 & 47.8 & {59.5$\pm$.6} & --- & {55.6}$\pm$.3 & \textbf{67.1$\pm$.3}\\
			$\texttt{REDDIT}$ (12K, 11 classes)      & {48.7}$\pm$.2 & ---  & {48.5$\pm$.5} & --- & {47.7}$\pm$.2 & \textbf{51.4$\pm$.2} \\
			$\texttt{COLLAB}$ (5K, 3 classes)         & {81.0}$\pm$.3 & 80.0 & --- & 83.6$\pm$.1 & 76.4$\pm$.4 & \textbf{88.3$\pm$.2}  \\
			$\texttt{IMDB-B}$ (1K, 2 classes)        & 71.9$\pm$1. & {73.6} & 75.1$\pm$1.1 & \textbf{76.9$\pm$3.6} & 71.2$\pm$.7 & 74.8$\pm$.3 \\
			$\texttt{IMDB-M}$ (1.5K, 3 classes)      & 47.7$\pm$.3 & \textbf{52.4} & 48.4$\pm$.5 & \textbf{52.8$\pm$4.6} & 48.8$\pm$.6 & 47.8$\pm$.7 \\
			\hline
		\end{tabular}
	}
	\caption{Mean accuracy and standard deviations for Large Social Network datasets.}
	\label{tab:res_social}
\end{table*}

As learning problems involving graph data are receiving a strong interest at the moment, consider a standard graph classification framework: $\Omega := (G, y) \in \mathcal{G} \times \mathcal{Y}$ is a finite family of graphs and available labels and one learns to map $\mathcal{G} \rightarrow \mathcal{Y}$.

Recently \cite{perslay} have introduced a powerful way of extracting topological information from graph structures. They make use of heat kernel signatures (HKS) for graphs \cite{hu2014stable}, a spectral family of signatures (with diffusion parameter $t > 0$) whose topological structure can be encoded in the extended persistence framework, yielding four types of topological features with exclusively finite persistence. We replicate their methodology, and on both HKS and extended persistence we refer to Sections 4.2 and 2 from \cite{perslay}. Schematically for diffusion time $t>0$ and graph $G(V, E)$ (with $V, E$ the sets of vertices and edges), the topological descriptors are computed as:
\begin{align}
\label{eq:graphpds} &\kappa_t := g \circ h,
\end{align}
where
\begin{align}
	&g : G(V, E) \in \mathcal{G} \xrightarrow[\text{signatures}]{\text{heat kernel}} \text{HKS}_{t}(G) \in \mathbb{R}^{|V|}, \\
	&h : \text{HKS}_{t}(G) \xrightarrow[\text{persistence}]{\text{extended}} \text{PD}(\text{HKS}_{t}(G)) \in \mathcal{D}^{4}.
\end{align}

For the entire set of problems to come we choose to use the same two HKS diffusion times to be $t_1=.1$ and $t_2=10$, so that Algorithm \ref{alg:atol} is used with topological map $\kappa := \kappa_{t_1} + \kappa_{t_2} : \mathcal{G} \rightarrow \mathcal{D}^8$, such that all in all 8 persistence diagrams are computed and considered per graph. For budget in Algorithm \ref{alg:atol} we choose $b=80$ for all experiments, which means Algorithm \ref{alg:atolfeat} will rely on approximating the mean measure on ten points per diagram type and HKS time filtration. We make no use of graph attributes on edges or vertices that some datasets do possess, and no other sort of features are collected, so that our results are solely based on the toplogical graph structure of the problems. To sum-up, Algorithm \ref{alg:atol} here simply consists in reducing the original problem from $\Omega$ to $\widetilde{\Omega} := (v_{\textsc{Atol}}\circ \kappa(G), y)$ with $v_{\textsc{Atol}}\circ \kappa(G) \in \mathbb{R}^{80}$. The embedding map $v_{\textsc{Atol}}$ from Algorithm ~\ref{alg:atolfeat} is computed using only 10\% of all diagrams from the training set, without supervision.

On each problem we perform a 10-fold cross-validation procedure and average the resulting accuracies; we report accuracies and standard deviations over ten such experiments. We use two sets of graph classification problems for benchmarking, one of Social Network origin and one of Chemoinformatics and Bioinformatics origin. They include small and large sets of graphs (\texttt{MUTAG} has 188 graphs, \texttt{REDDIT12K} has 12000), small and large graphs (\texttt{IMDB-M} has 13 nodes on average, \texttt{REDDIT5K} has more than 500), dense and sparse graphs (\texttt{FRANKENSTEIN} has around 12 edges per nodes, \texttt{COLLAB} has more than 2000), binary and multi-class problems (\texttt{REDDIT12K} has 11 classes), all available in the public repository \cite{benchmark}. Computations are run on a single laptop (i5-7440HQ 2.80 GHz CPU), in batch version for datasets smaller than a thousand observations and mini-batch version otherwise. Average computing time of Algorithm ~\ref{alg:atolfeat} (the average time to calibrate the vectorization map on the training set then compute the vectorization on the entire dataset), are: less than 1 second for datasets with less than a thousand observations, less than 5 seconds for datasets that have less than 5 thousand observations, 7.5 seconds for \texttt{REDDIT-5K}, and less than 16 seconds for the largest \texttt{REDDIT-12K} and the densest problem \texttt{COLLAB}.

We compare performances to the top scoring methods for these problems, to the best of our knowledge. Those methods are mostly graph kernel methods tailored to graph problems: two graph kernel methods based on random walks (RetGK1, RetGK11 from \cite{zhang2018retgk}), one graph embedding method based on spectral distances (FGSD from \cite{verma2017hunt}), two topological graph kernel method (WKPI-kM and WKPI-kC from \cite{qi2019}), one graph kernel combined with a graph neural network (GNTK from \cite{du19}). Finally PersLay from \cite{perslay} is a topological vectorization method learnt by a neural network that encodes most topological frameworks from the literature \--- landscapes, silhouettes, persistence images, etc. Note that the comparisons to PersLay were computed with the exact same persistence diagrams in most cases (except for a few cases where the authors used those same two HKS diffusion times then discarded one with no loss of performances) and the total budget for \textsc{Atol} ($b=80$) is a magnitude below that required to build the PersLay architecture (several hundreds nodes before the optimisation phase).
Competitor accuracy are quoted from their respective publication and should be interpreted as follows: for RetGK and WKPI and PersLay the evaluating procedure is done over ten 10-fold, just as ours is so the results directly compare; for FGSD the average accuracy over a single 10-fold is reported, and for GNTK the average accuracy and deviations is reported over a single 10-fold as well. When there are two or more methods under one label (e.g. RetGK1 and RetGK11), we always favorably reported the best outcome.

\begin{table*}
	\resizebox{.9\linewidth}{!}{ 
		\begin{tabular}{|l|rrrrr|r|}
			\hline
			\hspace{.8cm} method  & RetGK & FGSD & WKPI & GNTK & PersLay & \textsc{Atol} \\
			problem (size) & \cite{zhang2018retgk} & \cite{verma2017hunt} & \cite{qi2019} & \cite{du19} & \cite{perslay} & \\
			\hline
			$\texttt{MUTAG}$ (188)  & {90.3}$\pm$1.1 & \textbf{92.1} & 88.3$\pm$2.6 & 90.0$\pm$8.5 & 89.8$\pm$.9 & 88.3$\pm$.8\\
			$\texttt{COX2}$ (467)  & \textbf{81.4$\pm$.6} & --- & --- & --- & {80.9}$\pm$1. & 79.4$\pm$.7 \\
			$\texttt{DHFR}$ (756)  & {81.5$\pm$.9} & ---  & --- & --- & {80.3}$\pm$.8 & \textbf{{82.7}$\pm$.7}\\
			$\texttt{PROTEINS}$ (1113)  & \textbf{78.0$\pm$.3} & 73.4 & \textbf{78.5$\pm$.4} & 75.6$\pm$4.2 & {74.8}$\pm$.3 & 71.4$\pm$.6 \\
			$\texttt{NCI1}$ (4110)    & {84.5}$\pm$.2 & {79.8} & \textbf{87.5$\pm$.5} & 84.2$\pm$1.5 & 73.5$\pm$.3 & 78.5$\pm$.3 \\
			$\texttt{NCI109}$ (4127) & --- & 78.8 & \textbf{87.4$\pm$.3} & --- & 69.5$\pm$.3 & 77.0$\pm$.3 \\
			$\texttt{FRNKNSTN}$ (4337) & \textbf{76.4$\pm$.3} & --- & --- & --- & {70.7}$\pm$.4 & 72.9$\pm$.3\\
			\hline
		\end{tabular}
	}
	\caption{Mean accuracy and standard deviations for Chemoinformatics and Bioinformatics datasets, all binary classification problems.
	}
	\label{tab:res_bio}
\end{table*}

Our results Table \ref{tab:res_social} are {state-of-the-art} or substantially improving the {state-of-the-art} on the Large Social Network datasets that are difficult multi-class problems. The \texttt{REDDIT}s and \texttt{COLLAB} datasets all see major improvements in the mean accuracies, and those three datasets can readily be considered the most difficult problems (by size, graph density and number of classes) in the entire series. The results on the Chemoinformatics and Bioinformatics datasets Table \ref{tab:res_bio} are on par with or sometimes sub state-of-the-art, with a significant achievement on \texttt{DHFR}. It is not surprising that \textsc{Atol} is not always on par with the state-of-the-art, especially on the smaller binary classification datasets where considering the mean measure can potentially be too simple a model, easily refined upon \--- recall that contrary to competitors, \textsc{Atol} does not build a kernel or a neural network. Quantisation without supervision makes the learning process stricter, heavily dependent on the measure input: \textsc{Atol} is only capable of interpreting behavior with respect to the mean measure, therefore if some discriminant feature in a problem is found at a border of the measure space, as could be happening on the \texttt{PROTEINS} and \texttt{NCI}s datasets, it shall not be captured and there is no learning-room to change that. This is a liability, as well as a virtue of the method. We surmise that the \textsc{Atol} performances can be interpreted as a general optimal score for discriminative capacity with respect to the mean, in a problem. So for instance on the \texttt{IMDB} datasets, potentially whatever is gained on top of this baseline is obtained through supervision, at the cost of general discriminative power.

The {simplicity} and absence of tuning indicate robustness and generalisation power. Overall these results are especially positive seeing how Algorithm \ref{alg:atolfeat} has been employed with a universal configuration.

\subsection{A measured look at discrete dynamical systems}
\label{sec:ablation}

We now show the modularity capacity of the \textsc{Atol} framework, as well as its efficiency in compactly encoding information.

\begin{figure}[h]
	\centering
	\includegraphics[width= \columnwidth]{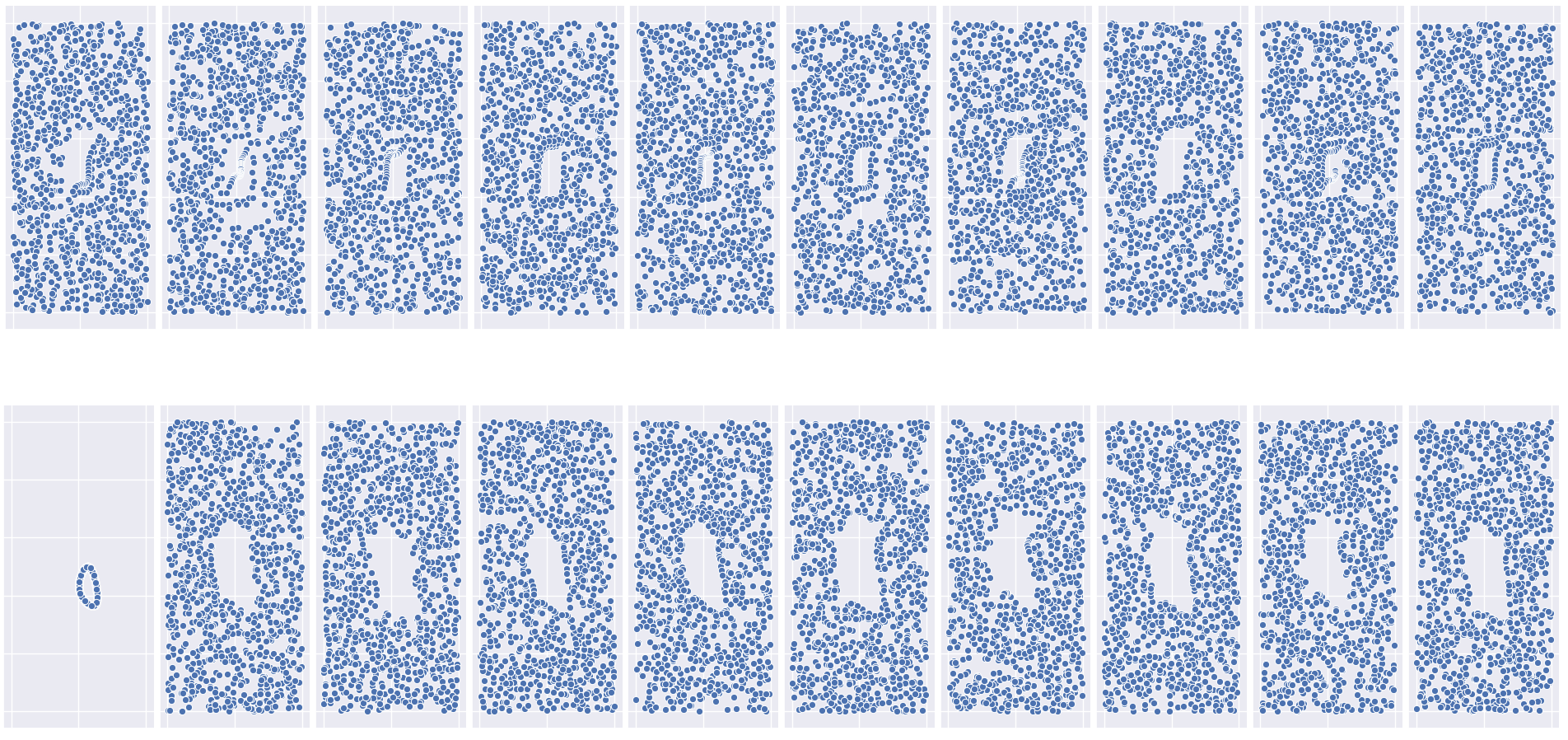}
	\caption{Synthetised orbits $x, y$ coordinates in $[0,1]^2$ for parameter $4.0$ (top) and $4.1$ (bottom).}
	\label{fig:orbits}
\end{figure}

\cite{Adams2017} use a synthetic, discrete dynamical system (used to model flows in DNA microarrays) with the following property: the resulting chaotic trajectories exhibit distinct topological characteristics depending on a parameter $r>0$. The dynamical system is: $x_{n+1} := x_n + r y_n (1-y_n) \mod 1$, and $y_{n+1} := y_n + r x_{n+1} (1-x_{n+1}) \mod 1$. With random initialisation and five different parameters $r \in \{2.5, 3.5, 4, 4.1, 4.3\}$, a thousand iterations per trajectory and a thousand orbits per parameter, a datasets of five thousand orbits is built. Figure~\ref{fig:orbits} shows a few orbits generated with parameters $r \in \{4.0, 4.1\}$. For orbits generated with parameter $r=4.1$, it happens that the initialisation spawns close to an attractor point that gives it the special shape as in the leftmost orbit. The problem of classifying this datasets according to their underlying parameter is rather uneasy and challenging. This dataset is commonly used for evaluating topological methods under the following experimental setup: a learning phase with a 70/30 split, and accuracy with standard deviation computed over a hundred such experiments. The
state-of-the-art accuracy of 87.7$\pm$1.0 with persistence diagrams is reported in \cite{perslay}.

Since those discrete orbits can be seen as measures in $[0,1]^2$, we instead decide to directly apply Algorithm \ref{alg:atol} on the observed point cloud, using the modularity of our framework \--- so in this instance $\kappa$ is the identity map. Therefore \textsc{Atol} is used here as a purely spatial approach and in this context, it is alike an image classification algorithm where instead of a fixed grid we have learnt center points to perform measurements. We present results in the form of a short ablation study Table \ref{tab:study} designed to illustrate influence of the small number of parameters from Algorithm \ref{alg:atolfeat}.

In this study we consider varying parameter $b \in \mathbb{N}^*$ for describing the measure space; replacing contrast functions $\Psi_i$ in Equation \eqref{contrast} with $\Phi_i(\cdot, \hat{\cb}_n) : x \mapsto \exp\big[-{\|x - c_i\|_2^2} / {\sigma^2_i(\hat{\cb}_n)} \big]$ for vectorization of the quantised space; and lastly changing the proportion of training observations used for deriving the quantization, with 10\% indicating that a random selection of a tenth of the measures from the training set were used to calibrate Algorithm \ref{alg:atolfeat}. We measure accuracies over 10 70/30 splits and for comparison purpose we also compute results for a 2D-grid quantization scheme labeled \texttt{grid}, that uses the same contrast family and a regular grid of size $\floor{\sqrt{b}} \times \floor{\sqrt{b}}$.

\begin{table*}[h]
	\begin{center}
		\resizebox{.9\linewidth}{!}{%
			\begin{tabular}{c|c|c|c|c||c|c||c|c||}
				\cline{2-9} &
				\multicolumn{4}{|c||}{\textbf{Budget}} & \multicolumn{2}{c||}{\textbf{Contrast functions}} & \multicolumn{2}{c||}{\textbf{Calibration}}
				\\
				\cline{2-9} &
				$b=4$ & $b=16$ & $b=36$ & \textcolor{blue}{${b = 100}$} & $\Phi$-Gaussian & \textcolor{blue}{$\Psi$-{Laplacian}} & \textcolor{blue}{\textbf{\texttt{10\%}}} & \texttt{100\%}
				\\ \hline
				\multicolumn{1}{|c|}{\textsc{Atol}} &
				56.3$\pm$1.6 & 83.1$\pm$2.2 & 89.6$\pm$1.3 & 93.8$\pm$.8 & 93.8$\pm$.5 & 93.8$\pm$.8 & 93.8$\pm$.8 & 93.6$\pm$.4				\\ 
				\multicolumn{1}{|c|}{} & 2.4 s & 3.1 s & 5.5 s & 12.7 s & 12.7 s & 12.7 s & 12.7 s & 50.2 s
				\\ \hline
				\multicolumn{1}{|c|}{\texttt{grid}} &
				55.8$\pm$1.1 & 82.7$\pm$.8 & 88.9$\pm$1.0 & 93.8$\pm$.7 & 94.2$\pm$.5 & 93.8$\pm$.7  & 93.8$\pm$.7 & 93.8$\pm$.7
				\\ \hline
		\end{tabular}}
	\end{center}
	\vskip-0.35cm
	\caption{Mean accuracy, deviation and vectorization time (including calibration step) over 10 experiments for \texttt{ORBIT5K}. Blue indicate parameters by default; only one parameter is varied at a time.}
	\label{tab:study}
\end{table*}

It is expected that a higher budget for vectorising the measure space will yield a better description of said space, and this intuition is confirmed by Table \ref{alg:atolfeat}. Although the differences are small, there is a slight advantage for operating \textsc{Atol} than a fixed \texttt{grid} at lower budgets, which is coherent with the intuition that the mean measure performs better than other procedures as a first approximation. Next, there does not seem to be significant differences from using Gaussian over Laplacian contrast functions on this experiment, although it can be the case on other problems. Understanding the ability of such contrast functions to describe some particular observation space is challenging and left for future work. Lastly, the percentage of observations used in the calibration part of the algorithm does not have a strong influence on the final result either (it does have a significant influence when the budget is lower then 80). This tells us that the calibration step is rather stable for a given level of information in a problem, and that our procedure is well-designed for dealing with problems online. Finally we report that when using budgets greater than 250 (i.e. finer than $12\times 12$ for the regular-grid), both methods reach comparable mean accuracies that are over 95\%. This indicates that this problem can be precisely described by a purely spatial approach, without topological descriptors.

\begin{figure}
	\centering
	\includegraphics[width=\columnwidth]{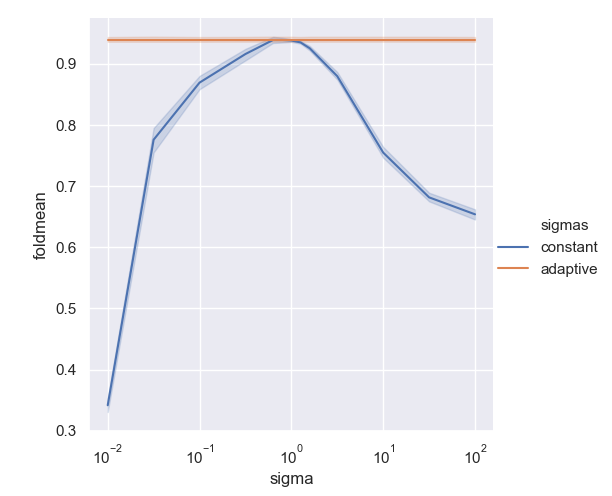}
	\caption{Classification accuracy and deviations for \texttt{Orbit5K} as $\sigma_1, \dots, \sigma_b = \sigma$ is varied (in blue), compared to the adaptive strategy of Equation \eqref{spread} (orange).}
	\label{fig:sigmas}
\end{figure}

Lastly, using the default parameters in Table \ref{tab:study} we compare the adaptive strategy of Equation \eqref{spread} for bandwiths $\sigma_1, \dots, \sigma_b$ to using identical constant values for those bandwiths. For investigating constant values we use the array $\mu \times 10^{[-2, -1.5, -1, -.5, -.2, -.1, 0, .1, .2, .5, 1, 1.5, 2]}$ where $\mu$ is the average distance between codebook points of $\hat{\cb}_n$. The results are shown Figure \ref{fig:sigmas}. This experiment shows that if a constant value for bandwiths can be found for optimal results, the adaptive strategy introduced in this paper already produces competitive results effortlessly.

\section{Conclusion}

This paper introduces an unsupervised vectorization method for measures in Euclidean spaces based on optimal quantization procedures, then shows how this method can be employed in machine learning contexts to exploit topological information. \textsc{Atol} is fast, has a simple design, is multifaceted and ties theoretical guarantees to practical efficiency.

From a practical viewpoint, we can guess that our method is less prone to bias and overfitting for two reasons: the centers are designed unsupervisedly (thus no possible overfitting) and the dimension of our vectorization is credibly low. Furthermore, effective insight can be gained as the method is interpretable: once the mean measure is computed one can observe its location (e.g. for diagrams, are centers close to the diagonal or not) and further derive information from it (e.g. in a classification task, center importance). For instance on a using a particular dataset of diagrams, one can learn if low persistence points are meaningful signal or not, with no preconceived hypothesis. Lastly \textsc{Atol} only depends on a simple parameter: the size $b$ of the codebook.



\bibliography{biblio}
\bibliographystyle{alpha}

\end{document}